\documentclass[reprint,aps,prl,superscriptaddress]{revtex4-1}

\usepackage{graphicx}
\usepackage{amsmath}
\usepackage{amssymb}
\usepackage{amsfonts}
\usepackage{color}
\usepackage{graphicx}
\usepackage{epstopdf}
\usepackage{braket}

\newcommand{\dd}[1]{\text{d}#1}
\newcommand{\vket}{\ket{\text{vac}}}

\begin{document}

\title{Parasitic Photon-Pair Suppression via Photonic Stop-Band Engineering}

\author{L. G. Helt}

\affiliation{Centre for Ultrahigh bandwidth Devices for Optical Systems (CUDOS), MQ Photonics Research Centre, QSciTech Research Centre, Department of Physics and Astronomy,
Macquarie University, NSW 2109, Australia}

\author{Agata M. Bra\'nczyk}

\affiliation{Perimeter Institute for Theoretical Physics, Waterloo, Ontario, N2L 2Y5, Canada}

\author{Marco Liscidini}

\affiliation{Dipartimento di Fisica, Universit\`{a} delgi Studi di Pavia, via Bassi 6, Pavia, Italy}

\author{M. J. Steel}

\affiliation{Centre for Ultrahigh bandwidth Devices for Optical Systems (CUDOS), MQ Photonics Research Centre, QSciTech Research Centre, Department of Physics and Astronomy,
Macquarie University, NSW 2109, Australia}

\begin{abstract}
We calculate that an appropriate modification of the field associated with only one of the photons of a photon pair can suppress generation of the pair entirely. From this general result, we develop a method for suppressing the generation of undesired photon pairs utilizing photonic stop bands.  For a third-order nonlinear optical source of frequency-degenerate photons, we calculate the modified frequency spectrum (joint spectral intensity) and show a significant increase in a standard metric, the coincidence to accidental ratio. These results open a new avenue for photon-pair frequency correlation engineering.
\end{abstract}

\maketitle

Nonlinear optics is often compounded by additional processes. In third-order nonlinear media these can include self- and cross-phase modulation~\cite{Kumar:2005}, various forms of four-wave mixing~\cite{Nodop:2009}, third-harmonic generation~\cite{Miyata:2011}, or solitonic phenomena~\cite{Redondo:2014}. While they can sometimes be made to work together, such as in supercontinuum generation~\cite{Zheltikov:2003}, in general many undesired processes may simultaneously compete unless care is taken to ensure that a single process dominates.  This is true of both classical and quantum processes~\cite{Helt:2013,Bell:2015}. 

In recent years spontaneous (quantum) nonlinear optical processes have emerged as a leading contender for the production of photons required for optical implementations of quantum information processing tasks~\cite{Kwiat:1995}. Such photons may enable secure communication~\cite{Shor:2000}, fundamental tests of bosonic interference~\cite{Gard:2015}, and exponential speedups of certain computations~\cite{Grover:2005}.  While second-order nonlinear media have often been used to produce the required indistinguishable photons~\cite{Ahlrichs:2016}, due to their efficiency and need for only a single incident laser, large-scale quantum optical technologies will likely be made possible with the scalability and integration capabilities provided by third-order nonlinear media such as silicon or silicon nitride~\cite{Gentry:2015}.  

\begin{figure}[hbt] 
\centering
\includegraphics[width=\columnwidth]{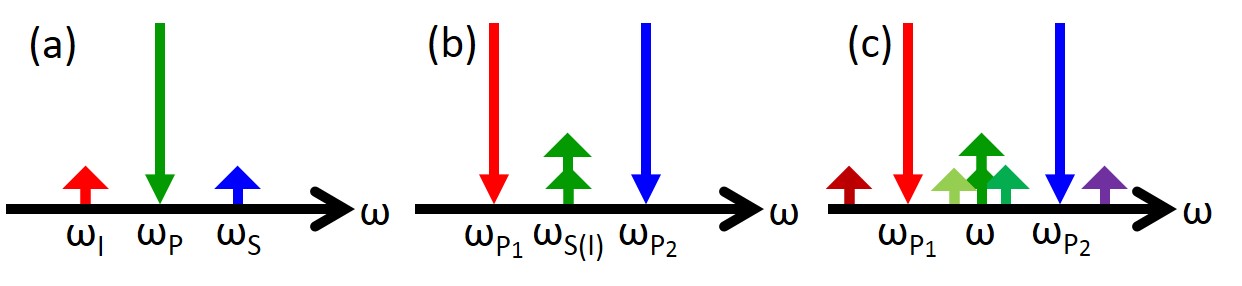}
\caption{Various SFWM processes with down and up arrows representing pump and generated frequencies, respectively: (a) single-pump, (b) idealized dual-pump, (c) dual-pump with undesired noise from parasitic single-pump processes.}
\label{fig:SFWM}
\end{figure}

The generation of non-classical states of light in these media occurs via spontaneous four-wave mixing (SFWM), a process that can be driven by either one or two pump beams. For a single incident pump narrowly centered at frequency $\omega_\text{P}$, single-pump SFWM stochastically generates signal and idler photons to either side of the pump frequency as dictated by energy conservation $\omega_\text{S}+\omega_\text{I}=2\omega_\text{P}$ [see Fig.~\ref{fig:SFWM}(a)]. In contrast, for two incident pumps narrowly centered at frequencies $\omega_{\text{P}_1}$ and $\omega_{\text{P}_2}$, dual-pump SFWM stochastically generates signal and idler photons located between the two pump frequencies according to $\omega_\text{S}+\omega_\text{I}=\omega_{\text{P}_1}+\omega_{\text{P}_2}$ [see Fig.~\ref{fig:SFWM}(b)]. 
Achieving simultaneous generation of spectrally degenerate photons with single-pump SFWM is experimentally challenging, for it requires simultaneously heralding a photon from each of two photon-pair sources. Yet while dual-pump SFWM can directly produce pairs of spectrally identical photons in a single device, it presents its own challenge in that it comes with parasitic noise photons due to each pump individually generating nonidentical pairs via single-pump SFWM~\cite{Silverstone:2014, He:2014} (see Fig.~\ref{fig:SFWM}(c)).

In this Letter we show that these parasitic photon pairs can be suppressed by appropriate engineering of photonic stop bands~\cite{Merklein:2015}. While previous studies have considered photonic crystal structures to enhance classical nonlinear optical processes~\cite{Mondia:2003, Becker:2006} or compensate for material dispersion in quantum processes~\cite{Corona:2007}, here we calculate that they may be used to suppress parasitic pair generation without compromising desired photon generation.  In particular, we show that this is possible because, unlike with a frequency filter, the frequency suppression of a photon can occur with a stop band placed at a \textit{different} frequency than that of a parasitic photon---namely, the frequency of its photon-pair partner. We focus here on the control of competing coherent photon generation processes, though note that, as with all photon-pair generation processes, additional (incoherent) noise sources of photons may also be present, e.g., Raman processes~\cite{Agrawal:2007}.

Employing the asymptotic fields formalism introduced in Ref.~\cite{Liscidini:2012}, we consider third-order nonlinear optical processes in an effectively one-dimensional structure, such as a waveguide or fiber, and therefore expand the electric field according to 
\begin{align}
\mathbf{E}\left(\mathbf{r}\right)=&\sum_a \int\dd{\omega}\,\sqrt{\frac{\hbar\omega_{a}}{4 \pi v_{a\omega}}}\varepsilon_{0}\nonumber \\
&\times\mathbf{e}_{a\omega}^{\text{in(out)}}\left(x,y\right)f_{a\omega}^{\text{in(out)}}\left(z\right)a_{a\omega}+\text{H.c.}
\end{align}
Here $\mathbf{e}_{a\omega}^{\text{in(out)}}\left(x,y\right)$ and $f_{a\omega}^{\text{in(out)}}\left(z\right)$ are, respectively, the transverse and longitudinal mode field distributions of the asymptotic-in(-out) modes labeled by $a$, normalized as in~\cite{Yang:2008}.  Additionally, $v_{a\omega}$ is the group velocity and the $a_{a\omega}$ are annihilation operators with commutation relations ${\left[a_{a\omega},a_{a^\prime \omega^\prime}^{\dag}\right]=\delta_{a a^\prime}\delta\left(\omega-\omega^\prime\right)}$. That being said, for simplicity, in all calculations that follow the fields involved will be assumed to occupy the same spatial mode (e.g. lowest-order linearly polarized, or LP, mode in a fiber), and we will use $a$ as a polarization label.

\begin{figure}[hbt] 
\centering
\includegraphics[width=\columnwidth]{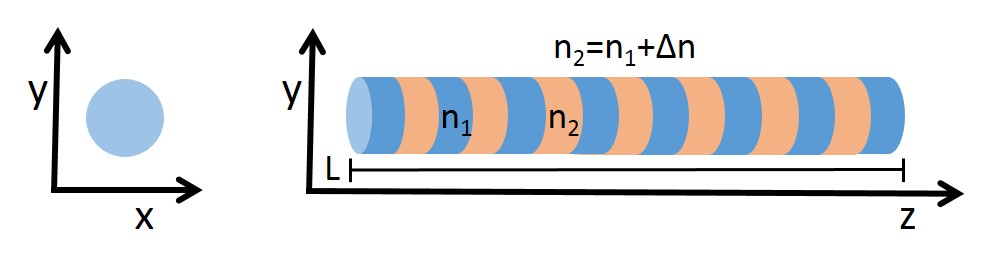}
\caption{Sketch of the class of structure with periodic index modulation that we consider for photon-pair generation.}
\label{fig:WG}
\end{figure}

We consider the production of deterministically separable photon pairs and so, for simplicity, assume a cubic or amorphous material and take the two pumps of the dual-pump process to be orthogonally polarized. In such a configuration the desired photon pairs are generated with orthogonal polarizations as well. The parasitic pairs from each pump, on the other hand, have the same polarization as the pump responsible for their generation. Using asymptotic fields to address this process allows us to leverage the stationary linear solutions of Maxwell's equations throughout a structure~\cite{Liscidini:2012}. Notably, the formalism straightforwardly generalizes the usual phase-matching function arising in a four-wave mixing interaction to 
\begin{align}
J_{ab}&\left(\omega_1,\omega_2,\omega_3,\omega_4\right)\nonumber \\
&=\frac{1}{\sqrt{\mathcal{N}_{ab}}}\int_0^L\dd{z}\,f_{a\omega_1}^{\text{in}}\left(z\right)f_{b\omega_2}^{\text{in}}\left(z\right)\left[f_{a\omega_3}^{\text{out}}\left(z\right)\right]^*\left[f_{b\omega_4}^{\text{out}}\left(z\right)\right]^*, \label{eq:J}
\end{align}
where $L$ is the length of the nonlinear structure, $a$ and $b$ can each take on the polarization labels $x$ and $y$, and $\mathcal{N}_{ab}$ is an appropriate normalization constant. In a straight waveguide with a constant refractive index in the longitudinal direction $f_{a\omega}^\text{in}\left(z\right)=f_{a\omega}^\text{out}\left(z\right)=e^{i k_a\left(\omega\right)z}$ and~\eqref{eq:J} reduces to a familiar sinc function~\cite{Alibart:2006}. In a straight waveguide with a periodic refractive index modulation in the longitudinal direction, the $f_{a\omega}^\text{in(out)}\left(z\right)$ correspond to Bloch modes. See Fig.~\ref{fig:WG} for a sketch of the type of structure we consider. Regardless of the linear properties of the waveguide,~\eqref{eq:J} shows that the product of these functions greatly influences the frequency spectrum of coherently generated photons.

In particular, when pumped with orthogonally polarized quasi-cw (classical) fields, in the limit of a low probability of pair production per pulse, the state of generated photons from a dual-pump SFWM process may be written
\begin{align}
\ket{\phi_{\text{gen}}}=&\frac{\zeta_{xx}}{\sqrt{2}}\int\dd{\omega}_{1}\dd{\omega}_{2}\,\phi_{xx}\left(\omega_{1},\omega_{2}\right)a_{x\omega_{1}}^{\dag}a_{x\omega_{2}}^{\dag}\vket \nonumber \\
&+\frac{\zeta_{yy}}{\sqrt{2}}\int\dd{\omega}_{1}\dd{\omega}_{2}\,\phi_{yy}\left(\omega_{1},\omega_{2}\right)a_{y\omega_{1}}^{\dag}a_{y\omega_{2}}^{\dag}\vket \nonumber \\
&+\zeta_{xy}\int\dd{\omega}_{1}\dd{\omega}_{2}\,\phi_{xy}\left(\omega_{1},\omega_{2}\right)a_{x\omega_{1}}^{\dag}a_{y\omega_{2}}^{\dag}\vket,
\end{align}
 where the joint spectral amplitudes
\begin{equation}
\phi_{ab}\left(\omega_1,\omega_2\right)=\alpha\left(\omega_1+\omega_2-\omega_a-\omega_b\right) J_{ab}\left(\omega_a,\omega_b,\omega_1,\omega_2\right), \label{eq:JSA}
\end{equation}
$\alpha\left(\omega\right)$ describes the (assumed identical) quasi-cw pump pulse spectral amplitudes, and the $\zeta_{ab}$ are set so as to normalize the joint spectral amplitudes according to $\int\dd{\omega_1}\dd{\omega_2}\left\vert\phi_{ab}\left(\omega_1,\omega_2\right)\right\vert^2=1$. Assuming a weak form birefringence of the waveguide or fiber that can be considered constant over the frequency range of interest and equal pump powers, they are related by $\zeta_{xy}=\frac{4}{3\sqrt{2}}\zeta_{xx}=\frac{4}{3\sqrt{2}}\zeta_{yy}$. Each term in the factor between the various $\zeta$ is due to a different effect: the 4 arises because there are four times as many ways four $E$-fields combine to generate cross-polarized photons than they do for a single co-polarized photon generation process, the 3 arises because the nonlinear tensor components of the underlying isotropic media satisfy~\cite{Boyd:2008} $\chi_3^{xxxx}=\chi_3^{yyyy}=\chi_3^{xyxy}/3$, and the $\sqrt{2}$ arises because of bosonic enhancement. The average number of pairs generated per pulse is then
\begin{align}
N_\text{pairs}=&\braket{\phi_\text{gen}|\phi_\text{gen}}\nonumber \\
\approx &\int\dd{\omega_1}\dd{\omega_2}\left(\left\vert\zeta_{xx}\right\vert^2\left\vert\phi_{xx}\left(\omega_1,\omega_2\right)\right\vert^2\right. \nonumber \\
&+\left.\left\vert\zeta_{yy}\right\vert^2\left\vert\phi_{yy}\left(\omega_1,\omega_2\right)\right\vert^2+\left\vert\zeta_{xy}\right\vert^2\left\vert\phi_{xy}\left(\omega_1,\omega_2\right)\right\vert^2\right)\nonumber \\
= &\left\vert\zeta_{xx}\right\vert^2+\left\vert\zeta_{yy}\right\vert^2+\left\vert\zeta_{xy}\right\vert^2=\frac{13}{4}\left\vert\zeta_{xy}\right\vert^2, \label{eq:pairs}
\end{align}
with the pair spectrum, or joint spectral intensity (JSI), given by 
\begin{align}
\Phi\left(\omega_1,\omega_2\right)=&\frac{\left\vert\bra{\phi_\text{gen}}\left(\frac{a_{x\omega_1}^{\dag}a_{x\omega_2}^{\dag}+a_{y\omega_1}^{\dag}a_{y\omega_2}^{\dag}}{\sqrt{2}}+a_{x\omega_1}^{\dag}a_{y\omega_2}^{\dag}\right)\vket\right\vert^2}{N_\text{pairs}}\nonumber \\
\approx &\frac{1}{26}\left(9\left\vert\phi_{xx}\left(\omega_1,\omega_2\right)\right\vert^2+9\left\vert\phi_{yy}\left(\omega_1,\omega_2\right)\right\vert^2\right.\nonumber \\
&+\left.8\left\vert\phi_{xy}\left(\omega_1,\omega_2\right)\right\vert^2\right), \label{eq:JSI}
\end{align}
where the approximation made in~\eqref{eq:pairs} and~\eqref{eq:JSI} is that the joint spectral amplitudes $\phi_{ab}\left(\omega_1,\omega_2\right)$ do not overlap due to the narrow frequency bandwidths associated with quasi-cw pump pulses. Here and throughout this letter we consider quasi-cw pump pulses to make the physics of the process as transparent as possible, but we note that the the method of calculation remains the same for all pump pulse durations.

It is clear from~\eqref{eq:pairs},~\eqref{eq:JSA}, and~\eqref{eq:J} that the efficiency of photon-pair generation via SFWM depends on the overlap of all \textit{four} transverse field distributions involved.  Indeed, as~\eqref{eq:J} involves coherent contributions from all creation points along the structure, properly engineering light propagation associated with $f_{a\omega_1}^\text{out}\left(z\right)$ in $J_{ab}\left(\omega_a,\omega_b,\omega_1,\omega_2\right)$ will affect photon generation not only at $\omega_1$, but at $\omega_2$ as well.  Returning to Figs.~\ref{fig:SFWM}(b) and~\ref{fig:SFWM}(c), we note that, while the desired degenerate photons are generated between $\omega_{\text{P}_1}$ and $\omega_{\text{P}_2}$, the generation of parasitic pairs includes frequencies outside this range. Thus, by properly engineering two photonic stop-bands, one centered at ${2\omega_{\text{P}_1}-\left(\omega_{\text{P}_2}+\omega_{\text{P}_1}\right)/2}$ and the other at ${2\omega_{\text{P}_2}-\left(\omega_{\text{P}_2}+\omega_{\text{P}_1}\right)/2}$, it is possible to inhibit the generation of parasitic photons without interfering with the creation of degenerate pairs. Referring back to~\eqref{eq:pairs}, this corresponds to reducing $\phi_{xx}\left(\omega_1,\omega_2\right)$ and $\phi_{yy}\left(\omega_1,\omega_2\right)$ over a desired collection bandwidth of $\phi_{xy}\left(\omega_1,\omega_2\right)$ by introducing stop-bands centered at $2\omega_x-\omega_d$ and $2\omega_y-\omega_d$, where $\omega_d\equiv\left(\omega_x+\omega_y\right)/2$, that, in turn, influence $f_{x\omega_1}^\text{out}\left(z\right)$ and $f_{y\omega_2}^\text{out}\left(z\right)$.

To introduce a stop-band in a straight waveguide we consider a periodic modulation of the refractive index in the longitudinal direction known as a Bragg grating. Its introduction opens a stop-band about a frequency $\omega_B$ with
\begin{equation}
f_{a\omega}^\text{out}\left(z\right)=G_{a\omega}^{+}\left(z\right)e^{i k_a\left(\omega_B\right) z}+G_{a\omega}^{-}\left(z\right)e^{-i k_a\left(\omega_B\right) z}, \label{eq:f_out}
\end{equation}
where
\begin{align}
G_{a\omega}^{+}\left(z\right)=&\frac{\left[\xi_{a\omega}\cosh\left(\xi_{a\omega}z\right)+i\delta_{a\omega}\sinh\left(\xi_{a\omega}z\right)\right]e^{i\delta_{a\omega}L}}{\xi_{a\omega}\cosh\left(\xi_{a\omega}L\right)+i\delta_{a\omega}\sinh\left(\xi_{a\omega}L\right)},\nonumber \\
G_{a\omega}^{-}\left(z\right)=&\frac{\kappa\sinh\left(\xi_{a\omega}z\right)e^{i\delta_{a\omega} L}}{i\xi_{a\omega}\cosh\left(\xi_{a\omega}L\right)-\delta_{a\omega}\sinh\left(\xi_{a\omega}L\right)},
\end{align}
with
$\xi_{a\omega}=\sqrt{\kappa^2-\delta_{a\omega}^2}$, $\delta_{a\omega}=k_a\left(\omega\right)-k_a\left(\omega_B\right)$, and ${\kappa=2\omega_B \Delta n/\left(\pi c\right)}$. Here $\omega_B$ is the center frequency of the Bragg grating, $\Delta n$ its index contrast, and ${\Lambda_a=\pi/k_a\left(\omega_B\right)}$ its period~\cite{Yariv:1973}. Similar expressions can be found for $f_{a\omega}^\text{in}\left(z\right)$~\cite{Supp}. For a grating that extends over the entire nonlinear region, approximating pumps as far detuned from grating resonances $f_{a\omega_a}^\text{in}\left(z\right)\approx e^{i k_a\left(\omega_a\right)z}$, and keeping only the first term of~\eqref{eq:f_out} on phase-matching grounds, the $J_{ab}$ of~\eqref{eq:JSA} is well-approximated by
\begin{align}
J_{ab}&\left(\omega_a,\omega_b,\omega_1,\omega_2\right)\nonumber \\
&=\frac{1}{\sqrt{\mathcal{N}_{ab}}}\int_0^L\dd{z}\,e^{i \left(\delta_{a\omega_a}+\delta_{b\omega_b}\right)z}\left[G_{a\omega_1}^{+}\left(z\right)\right]^*\left[G_{b\omega_2}^{+}\left(z\right)\right]^*,
\end{align}
with $\mathcal{N}_{ab}$ determined numerically. Thus, to suppress parasitic photon-pair generation centered about the degenerate dual-pump SFWM frequency $\omega_d$, due to the single-pump SFWM process driven by a quasi-cw pump with center frequency $\omega_a$, all that is necessary is to design a Bragg grating with a stop-band centered at ${\omega_{B_a}=2\omega_a-\omega_d}$. This will cause the product $\left[G_{a\omega_1}^{+}\left(z\right)\right]^*\left[G_{a\omega_2}^{+}\left(z\right)\right]^*$ of $J_{aa}$ to be greatly reduced over much of the range of $z$ near $\omega_{B_a}$, and $\omega_d$ as dictated by energy conservation, without affecting $J_{xy}$. For when $\delta_{a\omega}\approx 0$, with $\kappa L\gg 1$, $G_{a\omega}^{+}\left(z\right)\approx e^{-\kappa\left(L-z\right)}$. We note that, as opposed to parasitic-pair suppression with a standard Bragg grating, an alternate strategy would be to consider the enhancement of desired degenerate photons at $\omega_d$ due to a phase-shifted grating~\cite{Agrawal:1994}, but reserve its analysis for later work.

\begin{figure}[hbt] 
\centering
\includegraphics[width=\columnwidth]{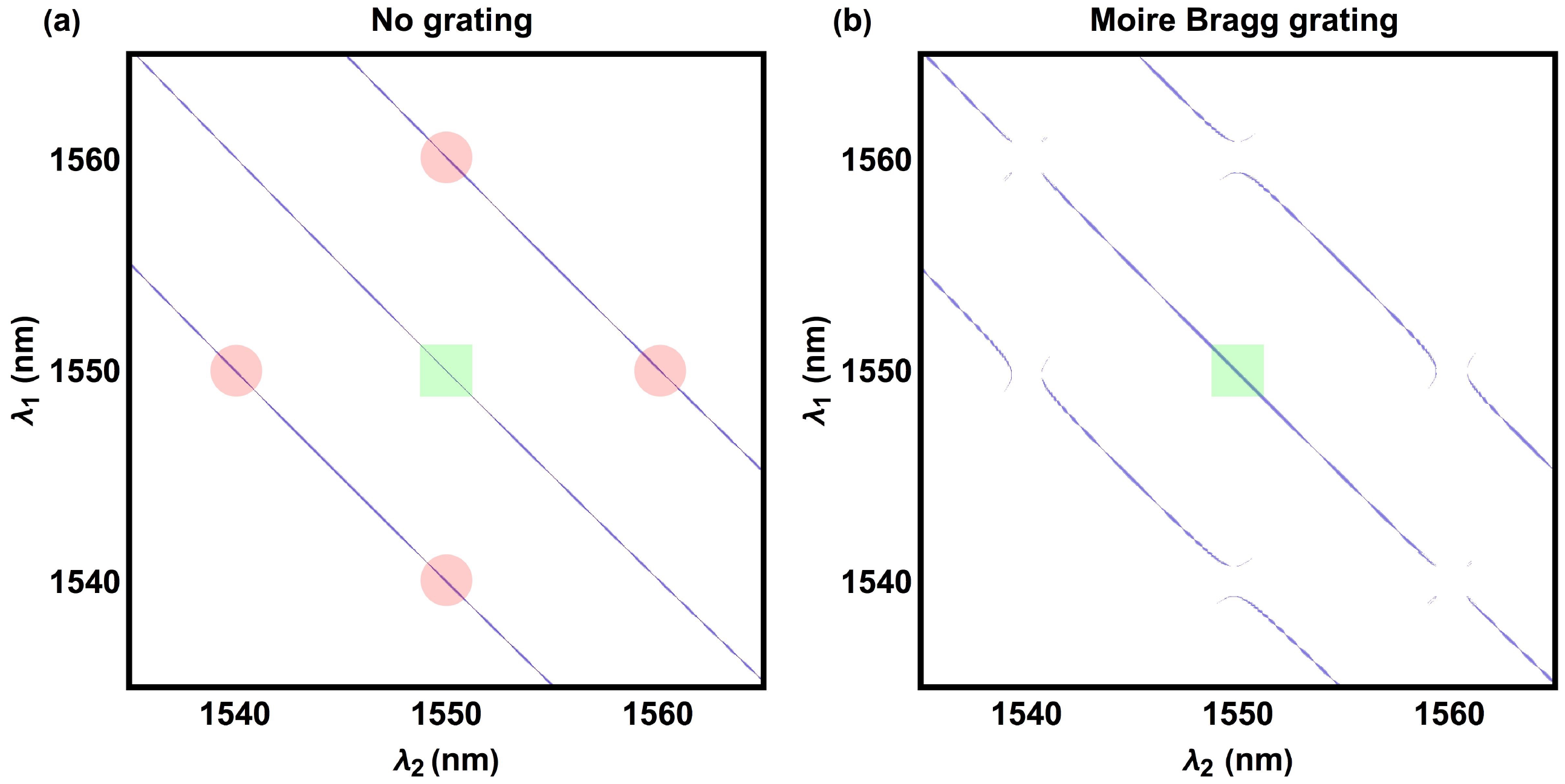}
\caption{Joint spectral intensities for our example system both a) without and b) with a moir\'{e} Bragg grating introduced. Desired degenerate photons are marked with a square and desired stop-bands with circles. The dual-pump process corresponds to the center diagonal bands, while parasitic single-pump x-polarized and y-polarized processes correspond to upper and lower diagonal bands, respectively.}
\label{fig:JSI}
\end{figure}

As a concrete example,  we consider a polarization maintaining fiber with a refractive index modeled by a standard Sellmeier equation~\cite{Malitson:1965}, and assume a linear birefringence between $x$ and $y$ polarizations of $3.3\times 10^{-5}$, conservative for such fibers~\cite{Corning}. In this system, we find that an $x$-polarized pump at $\omega_x=2\pi c/\left(1555~\text{nm}\right)$ and a $y$-polarized pump at $\omega_y=2\pi c/\left(1545~\text{nm}\right)$ are phase and energy matched to frequency degenerate photons produced via dual-pump SFWM at $\omega_d=2\pi c/\left(1550~\text{nm}\right)$.  Additionally, the pump at $\omega_x$ is phase and energy matched to parasitic pairs of photons produced via single-pump SFWM at $\omega_d$ and $2\pi c/\left(1560~\text{nm}\right)$ while the pump at $\omega_y$ is phase and energy matched to parasitic pairs of photons produced via single-pump SFWM at $\omega_d$ and $2\pi c/\left(1540~\text{nm}\right)$. For a fiber of length $L=50$~mm and intensity FWHM pump pulse durations of 10~ps, we plot the resultant JSI of \eqref{eq:JSI} in Fig.~\ref{fig:JSI}a. Note the three components of the plot corresponding to parasitic pairs from the $x$-polarized pump, parasitic pairs from the $y$-polarized pump, and desired dual-pump SFWM pairs from both pumps. To suppress these parasitic pairs we imagine introducing a dual stop-band, or moir\'{e}, grating~\cite{Reid:1990}, opening stop-bands centered at both $\omega_{B_x}=2\pi c/\left(1560~\text{nm}\right)$ and $\omega_{B_y}=2\pi c/\left(1540~\text{nm}\right)$ though reducing $\kappa$ by a factor of two. Assuming a grating index contrast of $\Delta n=2.1\times 10^{-3}$ and the grating to have the same effect on both polarizations, we calculate that 94,000 periods of length 536~nm ($L=50$~mm) can achieve a grating strength of $\kappa L\sim 137$ with photonic stop-bands of width $\Delta\lambda\approx\left(1550~\text{nm}\right)^2\kappa/\left[\pi n_{\text{SiO}_2}\left(1550~\text{nm}\right)\right]\approx 1.4$~nm centered at both $\omega_{B_x}$ and $\omega_{B_y}$~\cite{Erdogan:1997}. We note that SFWM has been observed in even shorter pieces of glass~\cite{Spring:2013,Yan:2015}. The JSI corresponding to the fiber with a grating introduced can be seen in Fig.~\ref{fig:JSI}b. Note how the components corresponding to parasitic pairs have been suppressed near $\omega_d=2\pi c/\left(1550~\text{nm}\right)$ compared to Fig.~\ref{fig:JSI}a.

\begin{figure}[hbt] 
\centering
\includegraphics[width=0.9\columnwidth]{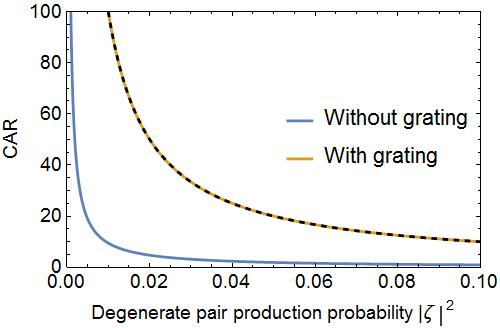}
\caption{Coincidence to accidental ratios for our example system both with and without the introduction of a moir\'{e} Bragg grating. The dashed line corresponds to the idealized maximum value of $1/\left\vert\zeta\right\vert^2$.}
\label{fig:CAR}
\end{figure}

A measurable effect of the introduction of the grating and suppression of parasitic photon-pairs should be a corresponding increase in the coincidence to accidental ratio (CAR)~\cite{Clark:2012}, a measure of the signal to noise of a photon-pair source. We examine this with a simple model for the CAR in which the average number of $a$-polarized photons collected per pulse, given efficiency $\eta_a$, average dark counts per pulse $d_a$, and a filter having spectral shape $h_a\left(\omega\right)$, is given by ${N_a=d_a+\eta_a\langle\int\dd{\omega}\,a_{a\omega}^{\dag}a_{a\omega}h_a\left(\omega\right)\rangle}$. Similarly, the average number of coincidence detections per pulse is given by ${N_{xy}=\eta_x\eta_y\langle\int\dd{\omega}\,a_{x\omega}^{\dag}a_{x\omega}h_x\left(\omega\right)\int\dd{\omega}\,a_{y\omega}^{\dag}a_{y\omega}h_y\left(\omega\right)\rangle}$ where the expectation values are taken with respect to $\ket{\phi_\text{gen}}$. Taking the filters to have the same spectral shape and both be centered at $\omega_d$, provided they are narrow enough that the $\left|\phi_{ab}\left(\omega_{1},\omega_{2}\right)\right|^{2}$ are essentially constant over their widths $\Delta$, we find
\begin{align}
N_{x}=&\eta_{x}\left|\zeta_{xy}\right|^{2}\Delta\int\text{d}\omega\left[\left|\phi_{xy}\left(\omega_{d},\omega\right)\right|^{2}+\frac{9}{4}\left|\phi_{xx}\left(\omega_{d},\omega\right)\right|^{2}\right]\nonumber \\
\approx &\eta_{x}\left|\zeta_{xy}\right|^{2}\Delta^2\left[\left|\phi_{xy}\left(\omega_{d},\omega_d\right)\right|^{2}+\frac{9}{4}\left|\phi_{xx}\left(\omega_{d},\omega_{B_x}\right)\right|^{2}\right],\nonumber \\
N_{y}=&\eta_{y}\left|\zeta_{xy}\right|^{2}\Delta\int\text{d}\omega\left[\left|\phi_{xy}\left(\omega,\omega_{d}\right)\right|^{2}+\frac{9}{4}\left|\phi_{yy}\left(\omega,\omega_{d}\right)\right|^{2}\right]\nonumber \\
\approx &\eta_{y}\left|\zeta_{xy}\right|^{2}\Delta^2\left[\left|\phi_{xy}\left(\omega_d,\omega_{d}\right)\right|^{2}+\frac{9}{4}\left|\phi_{yy}\left(\omega_{B_y},\omega_{d}\right)\right|^{2}\right],\nonumber \\
N_{xy}=&\eta_{x}\eta_{y}\left|\zeta_{xy}\right|^{2}\Delta^2\left|\phi_{xy}\left(\omega_{d},\omega_{d}\right)\right|^{2},
\end{align}
where we have made use of the symmetry ${\phi_{aa}\left(\omega_{1},\omega_{2}\right)=\phi_{aa}\left(\omega_{2},\omega_{1}\right)}$ and the narrow nature of $\phi_{ab}\left\vert\left(\omega_{1},\omega_{2}\right)\right\vert^2$ due to energy conservation. We have also neglected detector dark counts to focus on the effect of the grating. For the standard definition $CAR=N_{xy}/\left(N_{x}N_{y}\right)$ this leads to
\begin{equation}
CAR=\frac{1}{\left\vert\zeta\right\vert^2\left(1+\frac{9}{4}\frac{\left|\phi_{xx}\left(\omega_{d},\omega_{B_x}\right)\right|^{2}}{\left|\phi_{xy}\left(\omega_{d},\omega_{d}\right)\right|^{2}}\right)\left(1+\frac{9}{4}\frac{\left|\phi_{yy}\left(\omega_{B_y},\omega_{d}\right)\right|^{2}}{\left|\phi_{xy}\left(\omega_{d},\omega_{d}\right)\right|^{2}}\right)}, \label{eq:CAR}
\end{equation}
where we have defined $\left\vert\zeta\right\vert^2=\left|\zeta_{xy}\right|^{2}\Delta^2\left|\phi_{xy}\left(\omega_{d},\omega_{d}\right)\right|^{2}$. While the details of the CAR will depend on trade-offs between the grating width $\Delta\lambda$ and collection bandwidth $\Delta$, here we have made the simplifying assumption that parasitic photons can be suppressed over the collection bandwidth of interest $\Delta<2\pi c\Delta\lambda/\left(1550~\text{nm}\right)^2$. We note that much narrower collection bandwidths can be achieved with liquid crystal on silicon (LCoS) programmable optical processors~\cite{Finisar}. In Fig.~\ref{fig:CAR} we plot the CAR of~\eqref{eq:CAR} for our system both without and with the introduction of a moir\'{e} Bragg grating. For larger $\Delta$, the CAR associated with the grating structure will fall between these two limits.  Nonetheless, it is clear that the grating can greatly reduce the noise at 1550 nm associated with the parasitic single-pump processes, bringing the CAR towards the idealized maximum value of $1/\left|\zeta\right|^{2}$.

Throughout this work we have made simplifying choices in the interest of highlighting the underlying physics, but we stress that the main ideas are quite general.  For example, in addition to the ease of photon-pair separability offered by cross-polarized dual-pump SFWM,  photon pairs created via dual-pump SFWM could also be separated using parallel-polarized pumps with the nonlinear medium in a Sagnac loop and making use of so-called time-reversed Hong-Ou-Mandel interference~\cite{Chen:2007}. In place of a moir\'{e} grating in a fiber, in a strip waveguide one might imagine instead introducing different Bragg gratings along the right and left edges~\cite{Burla:2013}. Indeed, it is expected that similar results can be achieved in competing material platforms, such as silicon or a chalcogenide glass~\cite{Eggleton:2011}. In conclusion, Bragg gratings provide a new tool for photon-pair correlation engineering, allowing for competing undesirable processes to be suppressed. 

This research was supported in part by the ARC Centre for Ultrahigh bandwidth Devices for Optical Systems (CUDOS) (project number CE110001018), and by Perimeter Institute for Theoretical Physics. Research at Perimeter Institute is supported by the Government of Canada through Innovation, Science and Economic Development Canada and by the Province of Ontario through the Ministry of Research, Innovation and Science. The authors are grateful to Viola Introini for useful discussions.

\bibliography{bragg}

\begin{thebibliography}{38}%
\makeatletter
\providecommand \@ifxundefined [1]{%
 \@ifx{#1\undefined}
}%
\providecommand \@ifnum [1]{%
 \ifnum #1\expandafter \@firstoftwo
 \else \expandafter \@secondoftwo
 \fi
}%
\providecommand \@ifx [1]{%
 \ifx #1\expandafter \@firstoftwo
 \else \expandafter \@secondoftwo
 \fi
}%
\providecommand \natexlab [1]{#1}%
\providecommand \enquote  [1]{``#1''}%
\providecommand \bibnamefont  [1]{#1}%
\providecommand \bibfnamefont [1]{#1}%
\providecommand \citenamefont [1]{#1}%
\providecommand \href@noop [0]{\@secondoftwo}%
\providecommand \href [0]{\begingroup \@sanitize@url \@href}%
\providecommand \@href[1]{\@@startlink{#1}\@@href}%
\providecommand \@@href[1]{\endgroup#1\@@endlink}%
\providecommand \@sanitize@url [0]{\catcode `\\12\catcode `\$12\catcode
  `\&12\catcode `\#12\catcode `\^12\catcode `\_12\catcode `\%12\relax}%
\providecommand \@@startlink[1]{}%
\providecommand \@@endlink[0]{}%
\providecommand \url  [0]{\begingroup\@sanitize@url \@url }%
\providecommand \@url [1]{\endgroup\@href {#1}{\urlprefix }}%
\providecommand \urlprefix  [0]{URL }%
\providecommand \Eprint [0]{\href }%
\providecommand \doibase [0]{http://dx.doi.org/}%
\providecommand \selectlanguage [0]{\@gobble}%
\providecommand \bibinfo  [0]{\@secondoftwo}%
\providecommand \bibfield  [0]{\@secondoftwo}%
\providecommand \translation [1]{[#1]}%
\providecommand \BibitemOpen [0]{}%
\providecommand \bibitemStop [0]{}%
\providecommand \bibitemNoStop [0]{.\EOS\space}%
\providecommand \EOS [0]{\spacefactor3000\relax}%
\providecommand \BibitemShut  [1]{\csname bibitem#1\endcsname}%
\let\auto@bib@innerbib\@empty
\bibitem [{\citenamefont {Kumar}\ and\ \citenamefont
  {Yang}(2005)}]{Kumar:2005}%
  \BibitemOpen
  \bibfield  {author} {\bibinfo {author} {\bibfnamefont {S.}~\bibnamefont
  {Kumar}}\ and\ \bibinfo {author} {\bibfnamefont {D.}~\bibnamefont {Yang}},\
  }\href {http://jlt.osa.org/abstract.cfm?URI=jlt-23-6-2073} {\bibfield
  {journal} {\bibinfo  {journal} {J. Lightwave Technol.}\ }\textbf {\bibinfo
  {volume} {23}},\ \bibinfo {pages} {2073} (\bibinfo {year}
  {2005})}\BibitemShut {NoStop}%
\bibitem [{\citenamefont {Nodop}\ \emph {et~al.}(2009)\citenamefont {Nodop},
  \citenamefont {Jauregui}, \citenamefont {Schimpf}, \citenamefont {Limpert},\
  and\ \citenamefont {T\"{u}nnermann}}]{Nodop:2009}%
  \BibitemOpen
  \bibfield  {author} {\bibinfo {author} {\bibfnamefont {D.}~\bibnamefont
  {Nodop}}, \bibinfo {author} {\bibfnamefont {C.}~\bibnamefont {Jauregui}},
  \bibinfo {author} {\bibfnamefont {D.}~\bibnamefont {Schimpf}}, \bibinfo
  {author} {\bibfnamefont {J.}~\bibnamefont {Limpert}}, \ and\ \bibinfo
  {author} {\bibfnamefont {A.}~\bibnamefont {T\"{u}nnermann}},\ }\href
  {\doibase 10.1364/OL.34.003499} {\bibfield  {journal} {\bibinfo  {journal}
  {Opt. Lett.}\ }\textbf {\bibinfo {volume} {34}},\ \bibinfo {pages} {3499}
  (\bibinfo {year} {2009})}\BibitemShut {NoStop}%
\bibitem [{\citenamefont {Miyata}\ \emph {et~al.}(2011)\citenamefont {Miyata},
  \citenamefont {Petrov},\ and\ \citenamefont {Noack}}]{Miyata:2011}%
  \BibitemOpen
  \bibfield  {author} {\bibinfo {author} {\bibfnamefont {K.}~\bibnamefont
  {Miyata}}, \bibinfo {author} {\bibfnamefont {V.}~\bibnamefont {Petrov}}, \
  and\ \bibinfo {author} {\bibfnamefont {F.}~\bibnamefont {Noack}},\ }\href
  {\doibase 10.1364/OL.36.003627} {\bibfield  {journal} {\bibinfo  {journal}
  {Opt. Lett.}\ }\textbf {\bibinfo {volume} {36}},\ \bibinfo {pages} {3627}
  (\bibinfo {year} {2011})}\BibitemShut {NoStop}%
\bibitem [{\citenamefont {Blanco-Redondo}\ \emph {et~al.}(2014)\citenamefont
  {Blanco-Redondo}, \citenamefont {Husko}, \citenamefont {Eades}, \citenamefont
  {Zhang}, \citenamefont {Li}, \citenamefont {Krauss},\ and\ \citenamefont
  {Eggleton}}]{Redondo:2014}%
  \BibitemOpen
  \bibfield  {author} {\bibinfo {author} {\bibfnamefont {A.}~\bibnamefont
  {Blanco-Redondo}}, \bibinfo {author} {\bibfnamefont {C.}~\bibnamefont
  {Husko}}, \bibinfo {author} {\bibfnamefont {D.}~\bibnamefont {Eades}},
  \bibinfo {author} {\bibfnamefont {Y.}~\bibnamefont {Zhang}}, \bibinfo
  {author} {\bibfnamefont {J.}~\bibnamefont {Li}}, \bibinfo {author}
  {\bibfnamefont {T.}~\bibnamefont {Krauss}}, \ and\ \bibinfo {author}
  {\bibfnamefont {B.}~\bibnamefont {Eggleton}},\ }\href {\doibase
  doi:10.1038/ncomms4160} {\bibfield  {journal} {\bibinfo  {journal} {Nat.
  Commun.}\ }\textbf {\bibinfo {volume} {5}},\ \bibinfo {pages} {3160}
  (\bibinfo {year} {2014})}\BibitemShut {NoStop}%
\bibitem [{\citenamefont {Zheltikov}(2003)}]{Zheltikov:2003}%
  \BibitemOpen
  \bibfield  {author} {\bibinfo {author} {\bibfnamefont {A.}~\bibnamefont
  {Zheltikov}},\ }\href {\doibase 10.1007/s00340-003-1259-7} {\bibfield
  {journal} {\bibinfo  {journal} {Applied Physics B}\ }\textbf {\bibinfo
  {volume} {77}},\ \bibinfo {pages} {143} (\bibinfo {year} {2003})}\BibitemShut
  {NoStop}%
\bibitem [{\citenamefont {Helt}\ \emph {et~al.}(2013)\citenamefont {Helt},
  \citenamefont {Steel},\ and\ \citenamefont {Sipe}}]{Helt:2013}%
  \BibitemOpen
  \bibfield  {author} {\bibinfo {author} {\bibfnamefont {L.~G.}\ \bibnamefont
  {Helt}}, \bibinfo {author} {\bibfnamefont {M.~J.}\ \bibnamefont {Steel}}, \
  and\ \bibinfo {author} {\bibfnamefont {J.~E.}\ \bibnamefont {Sipe}},\ }\href
  {\doibase http://dx.doi.org/10.1063/1.4807503} {\bibfield  {journal}
  {\bibinfo  {journal} {Appl. Phys. Lett.}\ }\textbf {\bibinfo {volume}
  {102}},\ \bibinfo {pages} {201106} (\bibinfo {year} {2013})}\BibitemShut
  {NoStop}%
\bibitem [{\citenamefont {Bell}\ \emph {et~al.}(2015)\citenamefont {Bell},
  \citenamefont {McMillan}, \citenamefont {McCutcheon},\ and\ \citenamefont
  {Rarity}}]{Bell:2015}%
  \BibitemOpen
  \bibfield  {author} {\bibinfo {author} {\bibfnamefont {B.}~\bibnamefont
  {Bell}}, \bibinfo {author} {\bibfnamefont {A.}~\bibnamefont {McMillan}},
  \bibinfo {author} {\bibfnamefont {W.}~\bibnamefont {McCutcheon}}, \ and\
  \bibinfo {author} {\bibfnamefont {J.}~\bibnamefont {Rarity}},\ }\href
  {\doibase 10.1103/PhysRevA.92.053849} {\bibfield  {journal} {\bibinfo
  {journal} {Phys. Rev. A}\ }\textbf {\bibinfo {volume} {92}},\ \bibinfo
  {pages} {053849} (\bibinfo {year} {2015})}\BibitemShut {NoStop}%
\bibitem [{\citenamefont {Kwiat}\ \emph {et~al.}(1995)\citenamefont {Kwiat},
  \citenamefont {Mattle}, \citenamefont {Weinfurter}, \citenamefont
  {Zeilinger}, \citenamefont {Sergienko},\ and\ \citenamefont
  {Shih}}]{Kwiat:1995}%
  \BibitemOpen
  \bibfield  {author} {\bibinfo {author} {\bibfnamefont {P.~G.}\ \bibnamefont
  {Kwiat}}, \bibinfo {author} {\bibfnamefont {K.}~\bibnamefont {Mattle}},
  \bibinfo {author} {\bibfnamefont {H.}~\bibnamefont {Weinfurter}}, \bibinfo
  {author} {\bibfnamefont {A.}~\bibnamefont {Zeilinger}}, \bibinfo {author}
  {\bibfnamefont {A.~V.}\ \bibnamefont {Sergienko}}, \ and\ \bibinfo {author}
  {\bibfnamefont {Y.}~\bibnamefont {Shih}},\ }\href {\doibase
  10.1103/PhysRevLett.75.4337} {\bibfield  {journal} {\bibinfo  {journal}
  {Phys. Rev. Lett.}\ }\textbf {\bibinfo {volume} {75}},\ \bibinfo {pages}
  {4337} (\bibinfo {year} {1995})}\BibitemShut {NoStop}%
\bibitem [{\citenamefont {Shor}\ and\ \citenamefont
  {Preskill}(2000)}]{Shor:2000}%
  \BibitemOpen
  \bibfield  {author} {\bibinfo {author} {\bibfnamefont {P.~W.}\ \bibnamefont
  {Shor}}\ and\ \bibinfo {author} {\bibfnamefont {J.}~\bibnamefont
  {Preskill}},\ }\href {\doibase 10.1103/PhysRevLett.85.441} {\bibfield
  {journal} {\bibinfo  {journal} {Phys. Rev. Lett.}\ }\textbf {\bibinfo
  {volume} {85}},\ \bibinfo {pages} {441} (\bibinfo {year} {2000})}\BibitemShut
  {NoStop}%
\bibitem [{\citenamefont {Gard}\ \emph {et~al.}(2015)\citenamefont {Gard},
  \citenamefont {Motes}, \citenamefont {Olson}, \citenamefont {Rohde},\ and\
  \citenamefont {Dowling}}]{Gard:2015}%
  \BibitemOpen
  \bibfield  {author} {\bibinfo {author} {\bibfnamefont {B.~T.}\ \bibnamefont
  {Gard}}, \bibinfo {author} {\bibfnamefont {K.~R.}\ \bibnamefont {Motes}},
  \bibinfo {author} {\bibfnamefont {J.~P.}\ \bibnamefont {Olson}}, \bibinfo
  {author} {\bibfnamefont {P.~P.}\ \bibnamefont {Rohde}}, \ and\ \bibinfo
  {author} {\bibfnamefont {J.~P.}\ \bibnamefont {Dowling}},\ }\enquote
  {\bibinfo {title} {An introduction to boson-sampling},}\ in\ \href {\doibase
  10.1142/9789814678704_0008} {\emph {\bibinfo {booktitle} {From Atomic to
  Mesoscale}}}\ (\bibinfo  {publisher} {World Scientific},\ \bibinfo {year}
  {2015})\ Chap.\ \bibinfo {chapter} {Chapter 8}, pp.\ \bibinfo {pages}
  {167--192}\BibitemShut {NoStop}%
\bibitem [{\citenamefont {Grover}(2005)}]{Grover:2005}%
  \BibitemOpen
  \bibfield  {author} {\bibinfo {author} {\bibfnamefont {L.~K.}\ \bibnamefont
  {Grover}},\ }\href {\doibase 10.1103/PhysRevLett.95.150501} {\bibfield
  {journal} {\bibinfo  {journal} {Phys. Rev. Lett.}\ }\textbf {\bibinfo
  {volume} {95}},\ \bibinfo {pages} {150501} (\bibinfo {year}
  {2005})}\BibitemShut {NoStop}%
\bibitem [{\citenamefont {Ahlrichs}\ and\ \citenamefont
  {Benson}(2016)}]{Ahlrichs:2016}%
  \BibitemOpen
  \bibfield  {author} {\bibinfo {author} {\bibfnamefont {A.}~\bibnamefont
  {Ahlrichs}}\ and\ \bibinfo {author} {\bibfnamefont {O.}~\bibnamefont
  {Benson}},\ }\href {\doibase 10.1063/1.4939925} {\bibfield  {journal}
  {\bibinfo  {journal} {Appl. Phys. Lett.}\ }\textbf {\bibinfo {volume}
  {108}},\ \bibinfo {pages} {021111} (\bibinfo {year} {2016})}\BibitemShut
  {NoStop}%
\bibitem [{\citenamefont {Gentry}\ \emph {et~al.}(2015)\citenamefont {Gentry},
  \citenamefont {Shainline}, \citenamefont {Wade}, \citenamefont {Stevens},
  \citenamefont {Dyer}, \citenamefont {Zeng}, \citenamefont {Pavanello},
  \citenamefont {Gerrits}, \citenamefont {Nam}, \citenamefont {Mirin},\ and\
  \citenamefont {Popovi\'{c}}}]{Gentry:2015}%
  \BibitemOpen
  \bibfield  {author} {\bibinfo {author} {\bibfnamefont {C.~M.}\ \bibnamefont
  {Gentry}}, \bibinfo {author} {\bibfnamefont {J.~M.}\ \bibnamefont
  {Shainline}}, \bibinfo {author} {\bibfnamefont {M.~T.}\ \bibnamefont {Wade}},
  \bibinfo {author} {\bibfnamefont {M.~J.}\ \bibnamefont {Stevens}}, \bibinfo
  {author} {\bibfnamefont {S.~D.}\ \bibnamefont {Dyer}}, \bibinfo {author}
  {\bibfnamefont {X.}~\bibnamefont {Zeng}}, \bibinfo {author} {\bibfnamefont
  {F.}~\bibnamefont {Pavanello}}, \bibinfo {author} {\bibfnamefont
  {T.}~\bibnamefont {Gerrits}}, \bibinfo {author} {\bibfnamefont {S.~W.}\
  \bibnamefont {Nam}}, \bibinfo {author} {\bibfnamefont {R.~P.}\ \bibnamefont
  {Mirin}}, \ and\ \bibinfo {author} {\bibfnamefont {M.~A.}\ \bibnamefont
  {Popovi\'{c}}},\ }\href {\doibase 10.1364/OPTICA.2.001065} {\bibfield
  {journal} {\bibinfo  {journal} {Optica}\ }\textbf {\bibinfo {volume} {2}},\
  \bibinfo {pages} {1065} (\bibinfo {year} {2015})}\BibitemShut {NoStop}%
\bibitem [{\citenamefont {Silverstone}\ \emph {et~al.}(2014)\citenamefont
  {Silverstone}, \citenamefont {Bonneau}, \citenamefont {Ohira}, \citenamefont
  {Suzuki}, \citenamefont {Yoshida}, \citenamefont {Iizuka}, \citenamefont
  {Ezaki}, \citenamefont {Natarajan}, \citenamefont {Tanner}, \citenamefont
  {Hadfield}, \citenamefont {Zwiller}, \citenamefont {Marshall}, \citenamefont
  {Rarity}, \citenamefont {O'Brien},\ and\ \citenamefont
  {Thompson}}]{Silverstone:2014}%
  \BibitemOpen
  \bibfield  {author} {\bibinfo {author} {\bibfnamefont {J.~W.}\ \bibnamefont
  {Silverstone}}, \bibinfo {author} {\bibfnamefont {D.}~\bibnamefont
  {Bonneau}}, \bibinfo {author} {\bibfnamefont {K.}~\bibnamefont {Ohira}},
  \bibinfo {author} {\bibfnamefont {N.}~\bibnamefont {Suzuki}}, \bibinfo
  {author} {\bibfnamefont {H.}~\bibnamefont {Yoshida}}, \bibinfo {author}
  {\bibfnamefont {N.}~\bibnamefont {Iizuka}}, \bibinfo {author} {\bibfnamefont
  {M.}~\bibnamefont {Ezaki}}, \bibinfo {author} {\bibfnamefont {C.~M.}\
  \bibnamefont {Natarajan}}, \bibinfo {author} {\bibfnamefont {M.~G.}\
  \bibnamefont {Tanner}}, \bibinfo {author} {\bibfnamefont {R.~H.}\
  \bibnamefont {Hadfield}}, \bibinfo {author} {\bibfnamefont {V.}~\bibnamefont
  {Zwiller}}, \bibinfo {author} {\bibfnamefont {G.~D.}\ \bibnamefont
  {Marshall}}, \bibinfo {author} {\bibfnamefont {J.~G.}\ \bibnamefont
  {Rarity}}, \bibinfo {author} {\bibfnamefont {J.~L.}\ \bibnamefont {O'Brien}},
  \ and\ \bibinfo {author} {\bibfnamefont {M.~G.}\ \bibnamefont {Thompson}},\
  }\href {\doibase 10.1038/nphoton.2013.339} {\bibfield  {journal} {\bibinfo
  {journal} {Nat. Photon.}\ }\textbf {\bibinfo {volume} {8}},\ \bibinfo {pages}
  {105} (\bibinfo {year} {2014})}\BibitemShut {NoStop}%
\bibitem [{\citenamefont {He}\ \emph {et~al.}(2014)\citenamefont {He},
  \citenamefont {Clark}, \citenamefont {Collins}, \citenamefont {Li},
  \citenamefont {Krauss}, \citenamefont {Eggleton},\ and\ \citenamefont
  {Xiong}}]{He:2014}%
  \BibitemOpen
  \bibfield  {author} {\bibinfo {author} {\bibfnamefont {J.}~\bibnamefont
  {He}}, \bibinfo {author} {\bibfnamefont {A.~S.}\ \bibnamefont {Clark}},
  \bibinfo {author} {\bibfnamefont {M.~J.}\ \bibnamefont {Collins}}, \bibinfo
  {author} {\bibfnamefont {J.}~\bibnamefont {Li}}, \bibinfo {author}
  {\bibfnamefont {T.~F.}\ \bibnamefont {Krauss}}, \bibinfo {author}
  {\bibfnamefont {B.~J.}\ \bibnamefont {Eggleton}}, \ and\ \bibinfo {author}
  {\bibfnamefont {C.}~\bibnamefont {Xiong}},\ }\href {\doibase
  10.1364/OL.39.003575} {\bibfield  {journal} {\bibinfo  {journal} {Opt.
  Lett.}\ }\textbf {\bibinfo {volume} {39}},\ \bibinfo {pages} {3575} (\bibinfo
  {year} {2014})}\BibitemShut {NoStop}%
\bibitem [{\citenamefont {Merklein}\ \emph {et~al.}(2015)\citenamefont
  {Merklein}, \citenamefont {Kabakova}, \citenamefont {B\"{u}ttner},
  \citenamefont {Choi}, \citenamefont {Luther-Davies}, \citenamefont {Madden},\
  and\ \citenamefont {Eggleton}}]{Merklein:2015}%
  \BibitemOpen
  \bibfield  {author} {\bibinfo {author} {\bibfnamefont {M.}~\bibnamefont
  {Merklein}}, \bibinfo {author} {\bibfnamefont {I.~V.}\ \bibnamefont
  {Kabakova}}, \bibinfo {author} {\bibfnamefont {T.~F.~S.}\ \bibnamefont
  {B\"{u}ttner}}, \bibinfo {author} {\bibfnamefont {D.-Y.}\ \bibnamefont
  {Choi}}, \bibinfo {author} {\bibfnamefont {B.}~\bibnamefont {Luther-Davies}},
  \bibinfo {author} {\bibfnamefont {S.~J.}\ \bibnamefont {Madden}}, \ and\
  \bibinfo {author} {\bibfnamefont {B.~J.}\ \bibnamefont {Eggleton}},\ }\href
  {\doibase doi:10.1038/ncomms7396} {\bibfield  {journal} {\bibinfo  {journal}
  {Nat. Commun.}\ }\textbf {\bibinfo {volume} {6}},\ \bibinfo {pages} {6396}
  (\bibinfo {year} {2015})}\BibitemShut {NoStop}%
\bibitem [{\citenamefont {Mondia}\ \emph {et~al.}(2003)\citenamefont {Mondia},
  \citenamefont {van Driel}, \citenamefont {Jiang}, \citenamefont {Cowan},\
  and\ \citenamefont {Young}}]{Mondia:2003}%
  \BibitemOpen
  \bibfield  {author} {\bibinfo {author} {\bibfnamefont {J.~P.}\ \bibnamefont
  {Mondia}}, \bibinfo {author} {\bibfnamefont {H.~M.}\ \bibnamefont {van
  Driel}}, \bibinfo {author} {\bibfnamefont {W.}~\bibnamefont {Jiang}},
  \bibinfo {author} {\bibfnamefont {A.~R.}\ \bibnamefont {Cowan}}, \ and\
  \bibinfo {author} {\bibfnamefont {J.~F.}\ \bibnamefont {Young}},\ }\href
  {\doibase 10.1364/OL.28.002500} {\bibfield  {journal} {\bibinfo  {journal}
  {Opt. Lett.}\ }\textbf {\bibinfo {volume} {28}},\ \bibinfo {pages} {2500}
  (\bibinfo {year} {2003})}\BibitemShut {NoStop}%
\bibitem [{\citenamefont {Becker}\ \emph {et~al.}(2006)\citenamefont {Becker},
  \citenamefont {Wegener}, \citenamefont {Wong},\ and\ \citenamefont {von
  Freymann}}]{Becker:2006}%
  \BibitemOpen
  \bibfield  {author} {\bibinfo {author} {\bibfnamefont {C.}~\bibnamefont
  {Becker}}, \bibinfo {author} {\bibfnamefont {M.}~\bibnamefont {Wegener}},
  \bibinfo {author} {\bibfnamefont {S.}~\bibnamefont {Wong}}, \ and\ \bibinfo
  {author} {\bibfnamefont {G.}~\bibnamefont {von Freymann}},\ }\href {\doibase
  http://dx.doi.org/10.1063/1.2358295} {\bibfield  {journal} {\bibinfo
  {journal} {Appl. Phys. Lett.}\ }\textbf {\bibinfo {volume} {89}},\ \bibinfo
  {pages} {131122} (\bibinfo {year} {2006})}\BibitemShut {NoStop}%
\bibitem [{\citenamefont {Corona}\ and\ \citenamefont
  {U'Ren}(2007)}]{Corona:2007}%
  \BibitemOpen
  \bibfield  {author} {\bibinfo {author} {\bibfnamefont {M.}~\bibnamefont
  {Corona}}\ and\ \bibinfo {author} {\bibfnamefont {A.~B.}\ \bibnamefont
  {U'Ren}},\ }\href {\doibase 10.1103/PhysRevA.76.043829} {\bibfield  {journal}
  {\bibinfo  {journal} {Phys. Rev. A}\ }\textbf {\bibinfo {volume} {76}},\
  \bibinfo {pages} {043829} (\bibinfo {year} {2007})}\BibitemShut {NoStop}%
\bibitem [{\citenamefont {Agrawal}(2007)}]{Agrawal:2007}%
  \BibitemOpen
  \bibfield  {author} {\bibinfo {author} {\bibfnamefont {G.~P.}\ \bibnamefont
  {Agrawal}},\ }\href@noop {} {\emph {\bibinfo {title} {Nonlinear Fiber
  Optics}}},\ \bibinfo {edition} {4th}\ ed.\ (\bibinfo  {publisher} {Academic
  Press},\ \bibinfo {address} {Burlington, MA},\ \bibinfo {year}
  {2007})\BibitemShut {NoStop}%
\bibitem [{\citenamefont {Liscidini}\ \emph {et~al.}(2012)\citenamefont
  {Liscidini}, \citenamefont {Helt},\ and\ \citenamefont
  {Sipe}}]{Liscidini:2012}%
  \BibitemOpen
  \bibfield  {author} {\bibinfo {author} {\bibfnamefont {M.}~\bibnamefont
  {Liscidini}}, \bibinfo {author} {\bibfnamefont {L.~G.}\ \bibnamefont {Helt}},
  \ and\ \bibinfo {author} {\bibfnamefont {J.~E.}\ \bibnamefont {Sipe}},\
  }\href {\doibase 10.1103/PhysRevA.85.013833} {\bibfield  {journal} {\bibinfo
  {journal} {Phys. Rev. A}\ }\textbf {\bibinfo {volume} {85}},\ \bibinfo
  {pages} {013833} (\bibinfo {year} {2012})}\BibitemShut {NoStop}%
\bibitem [{\citenamefont {Yang}\ \emph {et~al.}(2008)\citenamefont {Yang},
  \citenamefont {Liscidini},\ and\ \citenamefont {Sipe}}]{Yang:2008}%
  \BibitemOpen
  \bibfield  {author} {\bibinfo {author} {\bibfnamefont {Z.}~\bibnamefont
  {Yang}}, \bibinfo {author} {\bibfnamefont {M.}~\bibnamefont {Liscidini}}, \
  and\ \bibinfo {author} {\bibfnamefont {J.~E.}\ \bibnamefont {Sipe}},\ }\href
  {\doibase 10.1103/PhysRevA.77.033808} {\bibfield  {journal} {\bibinfo
  {journal} {Phys. Rev. A}\ }\textbf {\bibinfo {volume} {77}},\ \bibinfo
  {pages} {033808} (\bibinfo {year} {2008})}\BibitemShut {NoStop}%
\bibitem [{\citenamefont {Alibart}\ \emph {et~al.}(2006)\citenamefont
  {Alibart}, \citenamefont {Fulconis}, \citenamefont {Wong}, \citenamefont
  {Murdoch}, \citenamefont {Wadsworth},\ and\ \citenamefont
  {Rarity}}]{Alibart:2006}%
  \BibitemOpen
  \bibfield  {author} {\bibinfo {author} {\bibfnamefont {O.}~\bibnamefont
  {Alibart}}, \bibinfo {author} {\bibfnamefont {J.}~\bibnamefont {Fulconis}},
  \bibinfo {author} {\bibfnamefont {G.~K.~L.}\ \bibnamefont {Wong}}, \bibinfo
  {author} {\bibfnamefont {S.~G.}\ \bibnamefont {Murdoch}}, \bibinfo {author}
  {\bibfnamefont {W.~J.}\ \bibnamefont {Wadsworth}}, \ and\ \bibinfo {author}
  {\bibfnamefont {J.~G.}\ \bibnamefont {Rarity}},\ }\href
  {http://stacks.iop.org/1367-2630/8/i=5/a=067} {\bibfield  {journal} {\bibinfo
   {journal} {New J. Phys.}\ }\textbf {\bibinfo {volume} {8}},\ \bibinfo
  {pages} {67} (\bibinfo {year} {2006})}\BibitemShut {NoStop}%
\bibitem [{\citenamefont {Boyd}(2008)}]{Boyd:2008}%
  \BibitemOpen
  \bibfield  {author} {\bibinfo {author} {\bibfnamefont {R.~W.}\ \bibnamefont
  {Boyd}},\ }\href@noop {} {\emph {\bibinfo {title} {Nonlinear Optics}}},\
  \bibinfo {edition} {3rd}\ ed.\ (\bibinfo  {publisher} {Academic Press},\
  \bibinfo {address} {Burlington, MA},\ \bibinfo {year} {2008})\BibitemShut
  {NoStop}%
\bibitem [{\citenamefont {Yariv}(1973)}]{Yariv:1973}%
  \BibitemOpen
  \bibfield  {author} {\bibinfo {author} {\bibfnamefont {A.}~\bibnamefont
  {Yariv}},\ }\href {\doibase 10.1109/JQE.1973.1077767} {\bibfield  {journal}
  {\bibinfo  {journal} {IEEE J. Quant. Electron.}\ }\textbf {\bibinfo {volume}
  {9}},\ \bibinfo {pages} {919} (\bibinfo {year} {1973})}\BibitemShut {NoStop}%
\bibitem [{Sup()}]{Supp}%
  \BibitemOpen
  \href@noop {} {}\bibinfo {note} {See Supplemental Material at
  \url{http://link.aps.org/supplemental/10.1103/PhysRevLett.118.073603} for
  details of the derivation of the longitudinal field distributions for
  asymptotic-in and-out fields in a Bragg grating.}\BibitemShut {Stop}%
\bibitem [{\citenamefont {Agrawal}\ and\ \citenamefont
  {Radic}(1994)}]{Agrawal:1994}%
  \BibitemOpen
  \bibfield  {author} {\bibinfo {author} {\bibfnamefont {G.~P.}\ \bibnamefont
  {Agrawal}}\ and\ \bibinfo {author} {\bibfnamefont {S.}~\bibnamefont
  {Radic}},\ }\href {\doibase 10.1109/68.313074} {\bibfield  {journal}
  {\bibinfo  {journal} {IEEE Photon. Tech. Lett.}\ }\textbf {\bibinfo {volume}
  {6}},\ \bibinfo {pages} {995} (\bibinfo {year} {1994})}\BibitemShut {NoStop}%
\bibitem [{\citenamefont {Malitson}(1965)}]{Malitson:1965}%
  \BibitemOpen
  \bibfield  {author} {\bibinfo {author} {\bibfnamefont {I.~H.}\ \bibnamefont
  {Malitson}},\ }\href@noop {} {\bibfield  {journal} {\bibinfo  {journal} {J.
  Opt. Soc. Am.}\ }\textbf {\bibinfo {volume} {55}},\ \bibinfo {pages} {1205}
  (\bibinfo {year} {1965})}\BibitemShut {NoStop}%
\bibitem [{Cor()}]{Corning}%
  \BibitemOpen
  \href@noop {} {}\bibinfo {howpublished}
  {\url{http://www.corning.com/media/worldwide/csm/documents/PANDA PM and RC
  PANDA Specialty Fiber.pdf}},\ \bibinfo {note} {accessed:
  2016-08-26}\BibitemShut {NoStop}%
\bibitem [{\citenamefont {Reid}\ \emph {et~al.}(1990)\citenamefont {Reid},
  \citenamefont {Ragdale}, \citenamefont {Bennion}, \citenamefont {Buus},\ and\
  \citenamefont {Stewart}}]{Reid:1990}%
  \BibitemOpen
  \bibfield  {author} {\bibinfo {author} {\bibfnamefont {D.~C.~J.}\
  \bibnamefont {Reid}}, \bibinfo {author} {\bibfnamefont {C.~M.}\ \bibnamefont
  {Ragdale}}, \bibinfo {author} {\bibfnamefont {I.}~\bibnamefont {Bennion}},
  \bibinfo {author} {\bibfnamefont {J.}~\bibnamefont {Buus}}, \ and\ \bibinfo
  {author} {\bibfnamefont {W.~J.}\ \bibnamefont {Stewart}},\ }\href {\doibase
  10.1049/el:19900007} {\bibfield  {journal} {\bibinfo  {journal} {Electron.
  Lett.}\ }\textbf {\bibinfo {volume} {26}},\ \bibinfo {pages} {10} (\bibinfo
  {year} {1990})}\BibitemShut {NoStop}%
\bibitem [{\citenamefont {Erdogan}(1997)}]{Erdogan:1997}%
  \BibitemOpen
  \bibfield  {author} {\bibinfo {author} {\bibfnamefont {T.}~\bibnamefont
  {Erdogan}},\ }\href {\doibase 10.1109/50.618322} {\bibfield  {journal}
  {\bibinfo  {journal} {J. Lightwave Technol.}\ }\textbf {\bibinfo {volume}
  {15}},\ \bibinfo {pages} {1277} (\bibinfo {year} {1997})}\BibitemShut
  {NoStop}%
\bibitem [{\citenamefont {Spring}\ \emph {et~al.}(2013)\citenamefont {Spring},
  \citenamefont {Salter}, \citenamefont {Metcalf}, \citenamefont {Humphreys},
  \citenamefont {Moore}, \citenamefont {Thomas-Peter}, \citenamefont
  {Barbieri}, \citenamefont {Jin}, \citenamefont {Langford}, \citenamefont
  {Kolthammer}, \citenamefont {Booth},\ and\ \citenamefont
  {Walmsley}}]{Spring:2013}%
  \BibitemOpen
  \bibfield  {author} {\bibinfo {author} {\bibfnamefont {J.~B.}\ \bibnamefont
  {Spring}}, \bibinfo {author} {\bibfnamefont {P.~S.}\ \bibnamefont {Salter}},
  \bibinfo {author} {\bibfnamefont {B.~J.}\ \bibnamefont {Metcalf}}, \bibinfo
  {author} {\bibfnamefont {P.~C.}\ \bibnamefont {Humphreys}}, \bibinfo {author}
  {\bibfnamefont {M.}~\bibnamefont {Moore}}, \bibinfo {author} {\bibfnamefont
  {N.}~\bibnamefont {Thomas-Peter}}, \bibinfo {author} {\bibfnamefont
  {M.}~\bibnamefont {Barbieri}}, \bibinfo {author} {\bibfnamefont {X.-M.}\
  \bibnamefont {Jin}}, \bibinfo {author} {\bibfnamefont {N.~K.}\ \bibnamefont
  {Langford}}, \bibinfo {author} {\bibfnamefont {W.~S.}\ \bibnamefont
  {Kolthammer}}, \bibinfo {author} {\bibfnamefont {M.~J.}\ \bibnamefont
  {Booth}}, \ and\ \bibinfo {author} {\bibfnamefont {I.~A.}\ \bibnamefont
  {Walmsley}},\ }\href {\doibase 10.1364/OE.21.013522} {\bibfield  {journal}
  {\bibinfo  {journal} {Opt. Express}\ }\textbf {\bibinfo {volume} {21}},\
  \bibinfo {pages} {13522} (\bibinfo {year} {2013})}\BibitemShut {NoStop}%
\bibitem [{\citenamefont {Yan}\ \emph {et~al.}(2015)\citenamefont {Yan},
  \citenamefont {Duan}, \citenamefont {Helt}, \citenamefont {Ams},
  \citenamefont {Withford},\ and\ \citenamefont {Steel}}]{Yan:2015}%
  \BibitemOpen
  \bibfield  {author} {\bibinfo {author} {\bibfnamefont {Z.}~\bibnamefont
  {Yan}}, \bibinfo {author} {\bibfnamefont {Y.}~\bibnamefont {Duan}}, \bibinfo
  {author} {\bibfnamefont {L.~G.}\ \bibnamefont {Helt}}, \bibinfo {author}
  {\bibfnamefont {M.}~\bibnamefont {Ams}}, \bibinfo {author} {\bibfnamefont
  {M.~J.}\ \bibnamefont {Withford}}, \ and\ \bibinfo {author} {\bibfnamefont
  {M.~J.}\ \bibnamefont {Steel}},\ }\href {\doibase
  http://dx.doi.org/10.1063/1.4937374} {\bibfield  {journal} {\bibinfo
  {journal} {Appl. Phys. Lett.}\ }\textbf {\bibinfo {volume} {107}},\ \bibinfo
  {pages} {231106} (\bibinfo {year} {2015})}\BibitemShut {NoStop}%
\bibitem [{\citenamefont {Clark}\ \emph {et~al.}(2012)\citenamefont {Clark},
  \citenamefont {Collins}, \citenamefont {Judge}, \citenamefont {M\"{a}gi},
  \citenamefont {Xiong},\ and\ \citenamefont {Eggleton}}]{Clark:2012}%
  \BibitemOpen
  \bibfield  {author} {\bibinfo {author} {\bibfnamefont {A.~S.}\ \bibnamefont
  {Clark}}, \bibinfo {author} {\bibfnamefont {M.~J.}\ \bibnamefont {Collins}},
  \bibinfo {author} {\bibfnamefont {A.~C.}\ \bibnamefont {Judge}}, \bibinfo
  {author} {\bibfnamefont {E.~C.}\ \bibnamefont {M\"{a}gi}}, \bibinfo {author}
  {\bibfnamefont {C.}~\bibnamefont {Xiong}}, \ and\ \bibinfo {author}
  {\bibfnamefont {B.~J.}\ \bibnamefont {Eggleton}},\ }\href {\doibase
  10.1364/OE.20.016807} {\bibfield  {journal} {\bibinfo  {journal} {Opt.
  Express}\ }\textbf {\bibinfo {volume} {20}},\ \bibinfo {pages} {16807}
  (\bibinfo {year} {2012})}\BibitemShut {NoStop}%
\bibitem [{Fin()}]{Finisar}%
  \BibitemOpen
  \href@noop {} {}\bibinfo {howpublished}
  {\url{http://www.finisar.com/optical-instrumentation}},\ \bibinfo {note}
  {accessed: 2016-08-24}\BibitemShut {NoStop}%
\bibitem [{\citenamefont {Chen}\ \emph {et~al.}(2007)\citenamefont {Chen},
  \citenamefont {Lee},\ and\ \citenamefont {Kumar}}]{Chen:2007}%
  \BibitemOpen
  \bibfield  {author} {\bibinfo {author} {\bibfnamefont {J.}~\bibnamefont
  {Chen}}, \bibinfo {author} {\bibfnamefont {K.~F.}\ \bibnamefont {Lee}}, \
  and\ \bibinfo {author} {\bibfnamefont {P.}~\bibnamefont {Kumar}},\ }\href
  {\doibase 10.1103/PhysRevA.76.031804} {\bibfield  {journal} {\bibinfo
  {journal} {Phys. Rev. A}\ }\textbf {\bibinfo {volume} {76}},\ \bibinfo
  {pages} {031804} (\bibinfo {year} {2007})}\BibitemShut {NoStop}%
\bibitem [{\citenamefont {Burla}\ \emph {et~al.}(2013)\citenamefont {Burla},
  \citenamefont {Cort\'{e}s}, \citenamefont {Li}, \citenamefont {Wang},
  \citenamefont {Chrostowski},\ and\ \citenamefont {Aza{\~n}a}}]{Burla:2013}%
  \BibitemOpen
  \bibfield  {author} {\bibinfo {author} {\bibfnamefont {M.}~\bibnamefont
  {Burla}}, \bibinfo {author} {\bibfnamefont {L.~R.}\ \bibnamefont
  {Cort\'{e}s}}, \bibinfo {author} {\bibfnamefont {M.}~\bibnamefont {Li}},
  \bibinfo {author} {\bibfnamefont {X.}~\bibnamefont {Wang}}, \bibinfo {author}
  {\bibfnamefont {L.}~\bibnamefont {Chrostowski}}, \ and\ \bibinfo {author}
  {\bibfnamefont {J.}~\bibnamefont {Aza{\~n}a}},\ }\href {\doibase
  10.1364/OE.21.025120} {\bibfield  {journal} {\bibinfo  {journal} {Opt.
  Express}\ }\textbf {\bibinfo {volume} {21}},\ \bibinfo {pages} {25120}
  (\bibinfo {year} {2013})}\BibitemShut {NoStop}%
\bibitem [{\citenamefont {Eggleton}\ \emph {et~al.}(2011)\citenamefont
  {Eggleton}, \citenamefont {Luther-Davies},\ and\ \citenamefont
  {Richardson}}]{Eggleton:2011}%
  \BibitemOpen
  \bibfield  {author} {\bibinfo {author} {\bibfnamefont {B.~J.}\ \bibnamefont
  {Eggleton}}, \bibinfo {author} {\bibfnamefont {B.}~\bibnamefont
  {Luther-Davies}}, \ and\ \bibinfo {author} {\bibfnamefont {K.}~\bibnamefont
  {Richardson}},\ }\href@noop {} {\bibfield  {journal} {\bibinfo  {journal}
  {Nat. Photon.}\ }\textbf {\bibinfo {volume} {5}},\ \bibinfo {pages} {141}
  (\bibinfo {year} {2011})}\BibitemShut {NoStop}%
\end{thebibliography}%

\end{document}